\documentclass[epjCONF]{svjour}
\usepackage{graphics}
\usepackage[varg]{txfonts} 
\usepackage[latin1]{inputenc}
\session-title{MESON2012 -- 12th International Workshop on Meson Production, Properties and Interaction}
\begin{document}
\title{Search for the $K_S \to 3\pi^0$ decay with the KLOE detector}
\author{M. Silarski\inst{1} on behalf of the KLOE -- 2 Collaboration
\thanks{
\email{Michal.Silarski@lnf.infn.it}\\
The KLOE -- 2 Collaboration: D.~Babusci, D.~Badoni, I.~Balwierz, G.~Bencivenni, C.~Bini, C.~Bloise, V.~Bocci, F.~Bossi, P.~Branchini, A.~Budano, 
L.~Caldeira~Balkest\aa hl, G.~Capon, F.~Ceradini, P.~Ciambrone, E.~Czerwi\'nski, E.~Dan\'e, E.~De~Lucia, G.~De~Robertis, A.~De~Santis,
A.~Di~Domenico, C.~Di~Donato, R.~Di~Salvo, D.~Domenici, O.~Erriquez, G.~Fanizzi, A.~Fantini, G.~Felici, S.~Fiore, P.~Franzini, P.~Gauzzi, G.~Giardina, S.~Giovannella, F.~Gonnella
E.~Graziani, F.~Happacher, B.~H\"oistad, L.~Iafolla, E.~Iarocci, M.~Jacewicz, T.~Johansson, A.~Kupsc, J.~Lee-Franzini, B.~Leverington, F.~Loddo,
S. Loffedo, G.~Mandaglio, M.~Martemianov, M.~Martini, M.~Mascolo, R.~Messi, S.~Miscetti, G.~Morello, D.~Moricciani, P.~Moskal, F.~Nguyen, A.~Passeri,
V.~Patera, I.~Prado~Longhi, A.~Ranieri, C.~F.~Redmer, P.~Santangelo, I.~Sarra, M.~Schioppa, B.~Sciascia, M.~Silarski, C.~Taccini, L.~Tortora,
G.~Venanzoni, R.~Versaci, W.~Wi\'slicki, M.~Wolke, J.~Zdebik
}}
\institute{\inst{1}Institute of Physics, Jagiellonian University, PL-30-059 Cracow, Poland}
\abstract{
The $K_S \to 3\pi^0$ decay is a pure CP violating process which, assuming
CPT invariance, allows one to investigate direct CP violation. 
This decay has not been observed so far, and the best upper limit on
the branching ratio $BR(K_S \to 3\pi^0)<1.2\cdot 10^{-7}$ is two orders
of magnitude larger than predictions based on the Standard Model.
In this article we present the search for the $K_S \to 3\pi^0$ decay performed
with the KLOE detector operating at the DA$\mathrm{\Phi}$NE $\phi$ -- factory at
the Frascati Laboratory.
We describe the analysis techniques used in the background rejection and
signal events selection, as well as the evaluation of almost five times lower
upper limit on the $K_S \to 3\pi^0$ branching ratio. We discuss shortly also
the perspectives for a new measurement using the KLOE -- 2 apparatus equipped
with a new inner tracker and the calorimeters at low $\theta$ angle. 
} 
\maketitle
\section{Introduction}
\label{intro}
The $\mathcal{CP}$ symmetry violation was discovered in 1964 by Christenson,
Cronin, Fitch and Turlay while studying the regeneration of the neutral $K$
mesons~\cite{Christensen}.
Since this unexpected discovery the $\mathcal{CP}$ violation parameters in the neutral kaon system
have been measured with a good precision by several experiments,
~and at present the main experimental effort is focused on studies
of the neutral $B$ and $D$ meson systems.
~However, there are still several interesting open issues in the kaon physics.
One of them is the $K_S \to 3\pi^0$ decay which, assuming the $\mathcal{CPT}$
invariance, allows one to investigate the direct $\mathcal{CP}$ symmetry
violation.
Despite several direct searches~\cite{snd,Matteo} and $K_S K_L$ interference
stu\-dies~\cite{cplear,na48b}, this decay remains undiscovered and the best
upper limit on the bran\-ching ratio $BR(K_S \to 3\pi^0)<1.2\cdot 10^{-7}$~\cite{Matteo,pdg2010}
is still two orders of magnitude larger than the predictions based on the
Standard Model: $BR(K_S \to 3\pi^0) \sim 2\cdot 10^{-9}$~\cite{Matteo}.\\
In this article we briefly describe the search of the $K_S \to 3\pi^0$ decay based on
the data sample gathered in 2004 -- 2005 with the KLOE detector operating at the $\phi$ -- factory
DA$\mathrm{\Phi}$NE of the Frascati Laboratory.
\section{$\mathcal{CP}$ violation in the neutral kaon system}
\label{sec:1}
From the point of view of strong interactions the $K^0$ meson is a particle with a corresponding
antiparticle $\overline{K}^{0}$. Violation of strangeness conservation by weak interaction allows for
transitions like $K^0 \to 2\pi \to \overline{K}^{0}$ or $K^0 \to 3\pi \to \overline{K}^{0}$. Thus,
the two strangeness eigenstates can oscillate one into another via the $\Delta S = 2$, second order
weak interactions.
Neutral kaons decay mainly to the two -- and three -- pion final states with a well defined
$\mathcal{CP}$ eigenvalues, therefore propagation and decays of these particles were
described in the basis of the $\mathcal{CP}$ operator eigenstates:
$
|K_{1}\rangle	= \frac{1}{\sqrt{2}} \left(|K^0\rangle + |\bar{K^0}\rangle\right)$ with CP~=~+1
and
$
|K_{2}\rangle	= \frac{1}{\sqrt{2}} \left(|K^0\rangle - |\bar{K^0}\rangle\right)$ with CP~=~-1.
$\mathcal{CP}$ conservation would imply that $|K_1\rangle$ state is allowed to decay only to two
pions with $\mathcal{CP}$~=~1,while the \textit{long} living
$|K_2\rangle$ decays only to three pions state with $\mathcal{CP}$~=~-1.\\
In 1964 an experiment by Christenson, Cronin, Fitch and Turlay, unexpectedly exhibited that the long -- lived kaon
can decay also to the two -- pion final states with branching ratio of about $2\cdot 10^{-3}$~\cite{Christensen}.
Thus, the neutral kaons states seen in nature are not the $\mathcal{CP}$ eigenstates defined before. However,
they still can be expressed in the ($|K_1\rangle$, $|K_2\rangle$) basis as:
$|K_{L}\rangle~=~\frac{1}{\sqrt{1+|\epsilon|^2}}~\left(|K_{2}\rangle~+~\epsilon|K_{1}\rangle\right)$~, and
$|K_{S}\rangle~=~\frac{1}{\sqrt{1+|\epsilon|^2}}~\left(|K_{1}\rangle~-~\epsilon|K_{2}\rangle\right)$~,
where $\epsilon$ express an admixture of a different $\mathcal{CP}$ eigenstate.
We can describe the $\mathcal{CP}$ symmetry breaking within the frame of two distinct mechanisms referred
to as \textit{direct} and \textit{indirect} breaking.
The \textit{indirect} violation corresponds to the statement that the eigenstates of both the electroweak
interactions are not exactly $\mathcal{CP}$
eigenstates but have small admixtures of the state with opposite $\mathcal{CP}$.
It is also possible, that $\mathcal{CP}$ violation occurs \textit{directly} in the weak decays themselves.\\
Typically the $\mathcal{CP}$ violation in the neutral kaon sector is characterized in terms of
the following amplitude ratios:
$\eta_{+-} = A( K_{L}\rightarrow \pi^{+}\pi^{-})/A(K_{S}\rightarrow \pi^{+}\pi^{-}) \cong \epsilon + \epsilon'$~,
and $\eta_{00} = A( K_{L}\rightarrow \pi^0\pi^0)/A(K_{S}\rightarrow \pi^0\pi^0) \cong \epsilon - 2\epsilon'$,
where the complex parameters $\epsilon'$ and $\epsilon$ express the \textit{direct} and \textit{indirect}
$\mathcal{CP}$ violation, respectively. In the framework of Standard Model an analogous CP breaking should
appear in the $K_S$ decays, for which we can define analogous amplitude ratios:
$\eta_{+-0} = A( K_{S}\rightarrow \pi^{+}\pi^{-} \pi^0)/A(K_{L}\rightarrow \pi^{+}\pi^{-}\pi^0) \cong \epsilon + \epsilon'_{+-0}$~,
and $\eta_{000} = A( K_{S}\rightarrow \pi^0\pi^0 \pi^0)/A(K_{L}\rightarrow \pi^0\pi^0 \pi^0) \cong \epsilon + \epsilon'_{000}$~.
As in the case of two -- pion decays the ratio contain direct $\mathcal{CP}$ violation parameters related
in the lowest order of the Chiral Perturbation Theory by the following equations:
$\epsilon'_{+-0} = \epsilon'_{000} = -2\epsilon'$~\cite{MPaver}.
While $\eta_{+-}$ and $\eta_{00}$ have been measured with a good precision,
the analogous parameters for $K_S$ are not well known~\cite{pdg2010}.
In particular, the $K_{S}\rightarrow \pi^0\pi^0\pi^0$ decay has been never observed,
and the branching ratio for this process is predicted to be very small in the Standard Model
(about $2\cdot 10^{-9}$). Therefore, studies of this decay demand high precision detectors
like KLOE  which will be briefly described in the next section.
\section{The KLOE experiment at DA$\mathrm{\Phi}$NE}
\label{KLOE:DAFNE}
\begin{figure*}
\centering
\resizebox{0.55\columnwidth}{!}{
\includegraphics{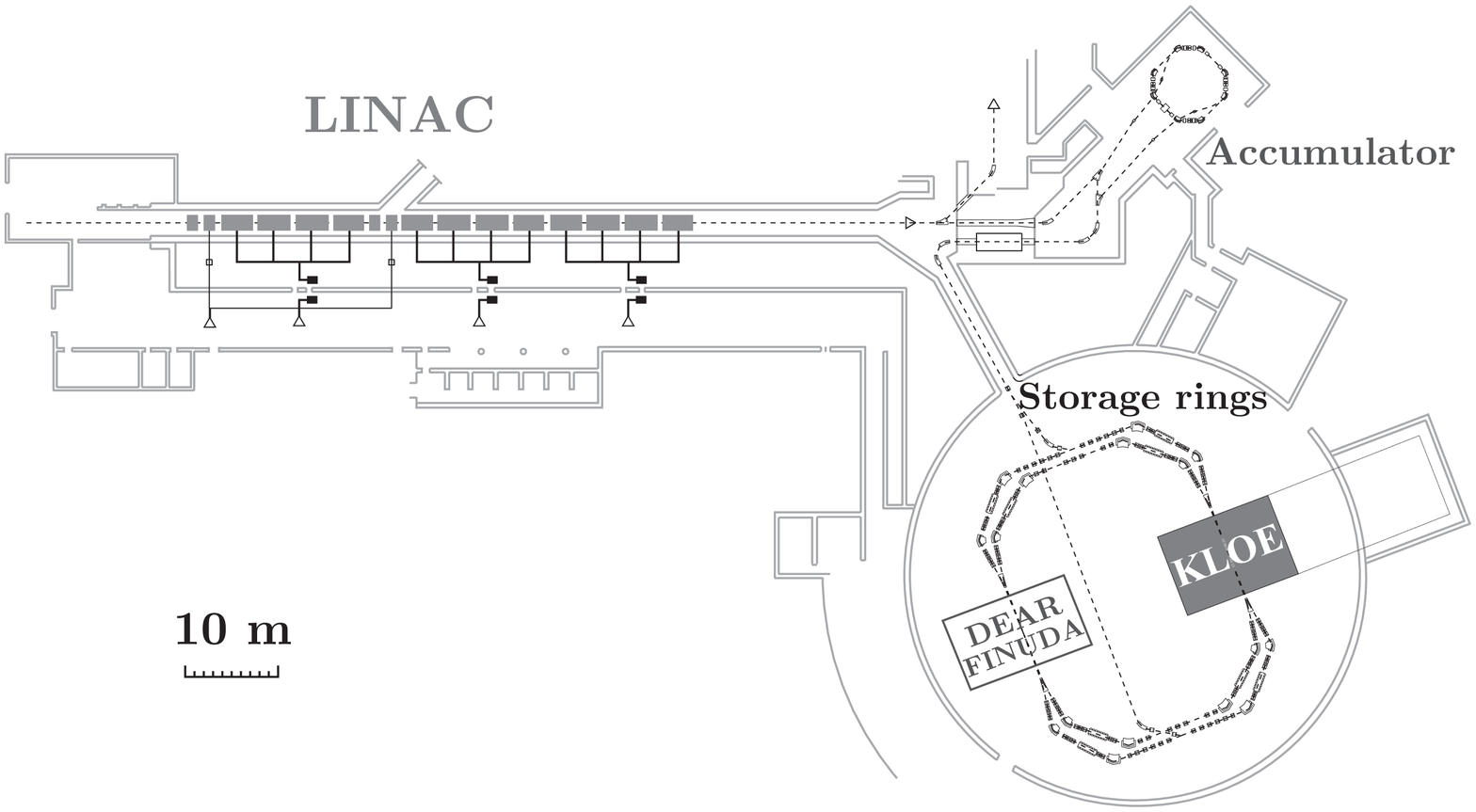}}
\resizebox{0.35\columnwidth}{!}{
\includegraphics{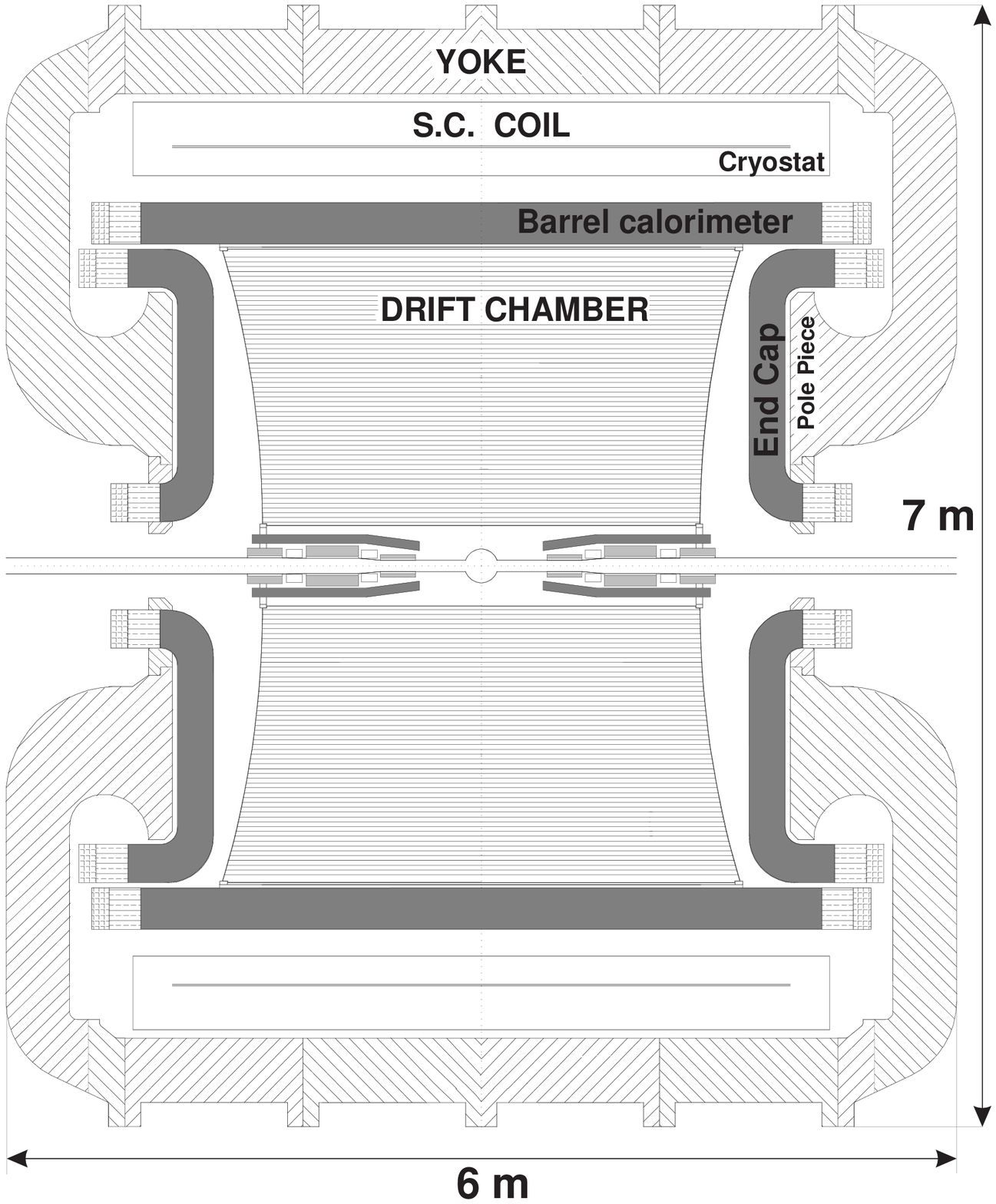}}
\caption{Left panel: Scheme of the DA$\mathrm{\Phi}$NE complex; Right panel:
Schematical view of the KLOE detector}
\label{dafne}
\end{figure*}

DA$\mathrm{\Phi}$NE is a $e^+e^-$ collider optimized to work with a center of mass energy around
the $\phi$ meson mass peak: $\sqrt{s}$ = 1019.45 MeV~\cite{kloe2008}. It consists of two rings
in which 120 bunches of both, electrons and positrons, are stored.
~Electrons are accelerated to final energy in the Linac
(see left panel of Fig.~\ref{dafne}),
stored and cooled in the accumulator, and then transferred to
a single bunch in the ring. Positrons instead, are created in an intermediate
station in the Linac, and then follow the same procedure as electrons.
The $e^+$ and $e^-$ beams collide with a small transverse momentum and produce 
$\phi$ mesons which are almost at rest ($\beta_{\phi} \approx 0.015$). These decay mainly to $K^+K^-$
(49\%), $K_SK_L$ (34\%), $\rho\pi$ (15\%) and $\eta\gamma$ (1.3\%)~\cite{pdg2010}.
The decay products are recorded using the KLOE detection setup, which is presented
schematically in the right panel of Fig.~\ref{dafne}. It consists of an about 3.3 long
cylindrical drift chamber with diameter equal to about 4 m, which is surrounded by  the electromagnetic
calorimeter. The detectors are placed in an axial magnetic field of superconducting
solenoid equal to $B~=~0.52$ T.\\
The KLOE tracking chamber was designed to detect all charged secondary products from the $K_L$
decay and measure their properties with great precision.
To minimize the $K_L$ regeneration, multiple Coulomb scattering and photon
absorption the chamber is constructed out of carbon fiber composite with low -- Z and low density,
and uses the gas mixture of helium~(90\%) and isobutane (10\%).
The KLOE drift chamber provides tracking in three dimensions with resolution in the bending
plane about 200 $\mu$m, resolution on the z-co\-or\-di\-na\-te measurement of about 2 mm and of 1 mm
on the decay vertex position. Momentum of the particle is determined from the curvature
of its trajectory in the magnetic field with a fractional accuracy $\sigma_p/p~=~0.4\%$
for polar angles larger than 45$^{\circ}$~\cite{kloe2008}.\\
The KLOE electromagnetic calorimeter consists
of a barrel built out of 24 trapezoidal shaped modules and two endcaps.
Each of the modules is built out of 1 mm scintillating fibers grouped in cells of 4.4x4.4 cm$^2$
and embedded in 0.5 mm lead foils, and it is read out from both sides by set of photomultipliers.
This detector allows
for measurements of particle energies and time with accuracies of 
$\sigma_E/E~=~5.7\%/\sqrt{E[GeV]}$ and $\sigma(t)~=~57 ps/\sqrt{E[GeV]}~\oplus~140~ps$, respectively.
Analysis of the signal amplitude distribution provides also the determination of the point where
the particle hit the calorimeter module with accuracy of about 1 cm in the plane transverse to
the fiber direction.
The longitudinal coordinate precision is energy dependent: $\sigma_z~=~1.2~cm/\sqrt{E[GeV]}$~\cite{kloe2008}.
\section{Search for the $K_S \to \pi^0\pi^0\pi^0$ decay}
At KLOE kaons arising from the $\phi$ decay move at low speed with their relative angle close to 180$^{\circ}$.
Therefore, observation of a $K_L$ ($K_S$ ) decay ensures the presence of a $K_S$ ($K_L$~)
meson travelling in the opposite direction.
The $K_S$ mesons are identified with high efficiency ($\sim$~34$\%$) via detection of the $K_L$ mesons
which cross the drift chamber
without decaying and then interact in the KLOE electromagnetic calorimeter (so called $K_S$ tag).
The $K_S$ four -- momentum vector is then determined using the measured position of the $K_L$ meson
and the known momentum of the $\phi$ meson, which is estimated as an average of the momentum distribution
measured using large angle $e^+e^-$ scattering.
The search for the $K_S\to 3\pi^0\to 6\gamma$ decay is then carried out by the selection of events
with six $\gamma$ quanta which momenta are reconstructed using time and energy measured by the electromagnetic
calorimeter.
Background for the searched decay originates mainly from the $K_S \to 2\pi^0$ events with two spurious
clusters from fragmentation of the electromagnetic showers (so called splitting) or accidental activity,
or from false $K_L$ identification for $\phi \to K_{S}K_{L} \to \pi^+\pi^-,3\pi^0$ events.
In the latter case charged pions from $K_S$ decays interact in the DA$\mathrm{\Phi}$NE low -- beta insertion quadrupoles,
ultimately simulating the $K_L$ interaction in the calorimeter, while $K_L$ decays close to the IP producing
six photons~\cite{Matteo}. To suppress this kind of background we first reject events
with charged particles coming from the vicinity of the interaction region. Moreover, we cut also on
the reconstructed velocity and energy of the tagging $K_L$ meson~\cite{silarskiPHD}.
In the next stage of the analysis we perform a kinematic fit with 11 constraints:
energy and momentum conservation, the kaon mass and the velocity of the six photons.
Cutting on the $\chi^2$ of the fit considerably reduces the background 
from bad quality reconstructed events with a very good signal efficiency.
In order to reject events with split and accidental clusters we look at the correlation between
two $\chi^2$ -- like discriminating variables $\chi^{2}_{2\pi}$ and $\chi^{2}_{3\pi}$. 
$\chi^{2}_{2\pi}$ is calculated by an algorithm selecting four out of six clusters best satisfying
the kinematic constraints of the two -- body decay, therefore it verifies the $K_S \to 2\pi^0 \to 4\gamma$
hypothesis. The pairing of clusters is based on the requirement $m_{\gamma\gamma} = m_{\pi^0}$,
and on the opening angle of the reconstructed pions trajectories
in the $K_S$ center of mass frame.
Moreover, we check the consistency of the energy and momentum conservation in the
$\phi \to K_S K_L, K_S \to 2\pi^0$ decay hypothesis~\cite{silarskiPHD}.
The $\chi^{2}_{3\pi}$ instead verifies the signal hypothesis by looking at
the reconstructed masses of three pions. For every choice of cluster pairs
we calculate the quadratic sum of the residuals between the nominal $\pi^0$
mass and the invariant masses of three photon pairs.
In order to improve the quality of the photon selection using $\chi^{2}_{2\pi}$,
we cut on the variable $\Delta E~=~(m_{\Phi}/2 - \sum E_{\gamma_{i}})/\sigma_{E}$
where $\gamma_i$ stands for the i--$th$ photon from  four chosen in the $\chi^{2}_{2\pi}$
estimator and $\sigma_E$ is the appropriate resolution. For $K_S \to 2\pi^0$ decays plus two background
clusters, we expect $\Delta E~\sim~$0, while for $K_S \to 3\pi^0$ $\Delta E~\sim~m_{\pi^0}/\sigma_E$.
At the end of the analysis we cut also on the minimal distance between photon clusters to refine rejection of events
with splitted clusters.\\
With preliminary cuts at the end of the analysis from 1.7 fb$^{-1}$ we
count 0 candidates with 0 background events expected
from Monte Carlo with an effective statistics of two times that of the data.
Hence, we have obtained the preliminary upper limit on the
$K_S \to 3\pi^0$ branching ratio at the 90$\%$ confidence level
$BR(K_S \to 3\pi^0) < 2.7 \cdot 10^{-8}$, which is almost five times
lower than the latest published result~\cite{Matteo}.
\section{Summary and outlook}
As a result of the full KLOE data set analysis, gathered in the 2004 -- 2005
data taking period, no events corresponding to the $K_S \to 3\pi^0$ decay
have been identified. Thus, we have set the upper limit for the $K_S \to 3\pi^0$
branching ratio at the 90$\%$ confidence level,
which is almost five times lower than the latest published result~\cite{Matteo}.
However, the search for the $K_S \to 3\pi^0$ decay will be continued
by the KLOE~--~2 collaboration~\cite{AmelinoCamelia}, which is continuing
and extending the physics program of its predecessor. 
For the forthcoming run the KLOE performance have been improved by adding new
subdetector systems: the tagger system for the $\gamma\gamma$ physics,
the Inner Tracker based on the Cylindrical GEM technology and two calorimeters
in the final focusing region~\cite{Moricciani:2012zza}. These new calorimeters
will increase the acceptance of the detector, while the new inner detector for
the determination of the $K_S$ vertex will significantly
reduce the contribution of the background processes involving charged particles.
Increasing the statistics and acceptance of the detector while significantly reducing the background gives
the realistic chances to observe the $K_S \to 3\pi^0$ decay for the first time in the near future.
\begin{acknowledgement}
We acknowledge the support by the Polish National Science Center and by the Foundation for Polish Science.
\end{acknowledgement}

\end{document}